\documentclass[referee]{raa}            
\usepackage{graphicx,times}             
\usepackage{lineno}
\usepackage[caption=false]{subfig}
\usepackage{multirow}
\usepackage{graphicx}
\usepackage{booktabs}
\usepackage{natbib}
\usepackage{amsmath,amssymb}
\usepackage[T1]{fontenc}
\begin{document}

   \title{Highly Variable $\gamma$-Ray Emission of CTD 135 and Implications for its Compact Symmetric Structure}

   \volnopage{Vol.0 (200x) No.0, 000--000}      
   \setcounter{page}{1}          

   \author{Ying-Ying Gan
      \inst{1}
   \and Hai-Ming Zhang
      \inst{2}
   \and Jin Zhang
      \inst{3\dag}
   \and Xing Yang
      \inst{1}
   \and Ting-Feng Yi
      \inst{4}
   \and Yun-Feng Liang
      \inst{1}
   \and En-Wei Liang
      \inst{1}
   }

   \institute{Guangxi Key Laboratory for Relativistic Astrophysics, School of Physical Science and Technology, Guangxi University, Nanning 530004, People's Republic of China
     \and
         School of Astronomy and Space Science, Nanjing University, Nanjing 210023, People's Republic of China
     \and
         School of Physics, Beijing Institute of Technology, Beijing 100081, People's Republic of China; j.zhang@bit.edu.cn
     \and
         Department of Physics, Yunnan Normal University, Kunming 650500, People's Republic of China
   }

   \date{Received~~201X month day; accepted~~201X~~month day}

\abstract{The $\gamma$-ray emission properties of CTD 135, a typical compact symmetric object (CSO), are investigated with $\sim$11-year Fermi/LAT observations. We show that it has bright and significantly variable GeV emission, with the $\gamma$-ray luminosity of $L_{\gamma}\sim10^{47}$ erg s$^{-1}$ and a variation index of TS$_{\rm var}=1002$. A quasi-periodic oscillation (QPO) with a periodicity of $\sim$460 days is detected in the global 95\% false-alarm level. These $\gamma$-ray emission features are similar to that of blazars. Its broadband spectral energy distribution (SED) can be attributed to the radiations of the relativistic electrons accelerated in the core region and the extended region. The SED modeling shows that the $\gamma$-rays are from the core region, which has a Doppler boosting factor of $\delta\sim10.8$ and relativistically moves with a small viewing angle, being similar to blazar jets. On the base of the analysis results, we propose that the episodic activity of the central engine in CTD 135 results in a blazar-like jet and the bubble-like lobes as the Fermi bubbles in the Galaxy. The strong $\gamma$-ray emission with obvious variability is from the jet radiations and the symmetric radio structure is attributed to the bubbles. The jet radiation power and disk luminosity in units of Eddington luminosity of CTD 135 follow the same relation as other young radio galaxies, indicating that its jet radiation may be also driven by the Eddington ratio.
\keywords{galaxies: active---galaxies: jets---radio continuum: galaxies---gamma rays: galaxies}
}

   \authorrunning{Ying-Ying Gan et al. }            
   \titlerunning{$\gamma$-ray Emitting CSO CTD 135 }  

   \maketitle
%
%
\section{Introduction}           
\label{sect:intro}

Compact symmetric objects (CSOs) are of a sub-class of misaligned active galactic nuclei (AGNs), being characterized by a symmetric and sub-kiloparsec (sub-kpc) radio structure (\citealt{1980ApJ...236...89P, 1994ApJ...432L..87W, 1996ApJ...460..634R}). The radio spectra of some CSOs exhibit a turnover around a few GHz frequency (\citealt{1997A&A...321..105D, 2003PASA...20...19M}), similar to gigahertz peaked-spectrum sources (GPSs, \citealt{1998PASP..110..493O, 2009AN....330..120F}). The radio emission of CSOs displays weak variability (\citealt{2001AJ....122.1661F}) and low polarization (\citealt{2000ApJ...534...90P}). Although they may also be confined to the sub-kpc scale by a dense and turbulent ambient medium (\citealt{1984AJ.....89....5V, 1998PASP..110..493O}), it is still lack of observational evidence that the amount of ambient gas can supply enough confinement. They are generally thought as young radio galaxies and the progenitors of the extended radio galaxies (e.g., \citealt{1982A&A...106...21P, 1995A&A...302..317F, 1996ApJ...460..612R}). However, only a fraction of CSOs may evolve into large-scale radio galaxies (\citealt{2012ApJ...760...77A}). The radio luminosity would decrease by an order of magnitude as the radio size of source grows from a few kpc to hundreds of kpc (\citealt{1995A&A...302..317F, 1996cyga.book..209B, 1996ApJ...460..634R,  1996ApJ...460..612R}), and the compact sources will evolve into objects less powerful than the most powerful classical doubles (\citealt{1997AJ....113..148O, 1998A&AS..131..303S}). As representatives of young AGNs, CSOs are of interest for understanding the early evolution and the jet radiation physics of AGNs.

Strong $\gamma$-ray emission from the parsec-scale (pc-scale) lobes of young radio sources was theoretically predicted (\citealt{2008ApJ...680..911S, 2007MNRAS.376.1630K, 2009MNRAS.395L..43K, 2011MNRAS.412L..20K, 2014ApJ...780..165M}), but systematic search for CSOs in the $\gamma$-ray band is rather unsuccessful  (\citealt{2016AN....337...59D}). So far, only a few CSOs were detected in the $\gamma$-ray band, including PMN J1603--4904 (\citealt{2014A&A...562A...4M, 2015A&A...574A.117M}), PKS 1718--649 (\citealt{2016ApJ...821L..31M}), NGC 3894 (\citealt{2020A&A...635A.185P}), and TXS 0128+554 (\citealt{2020ApJ...899..141L}).
The origin of $\gamma$-ray emission in CSOs is still debated. The $\gamma$-ray emission of TXS 0128+554 likely originates in the inner jet/core region since the predicted flux by the inverse Compton emitting cocoon model is seriously lower than the observed Fermi/LAT flux (\citealt{2020ApJ...899..141L}). Thus, the $\gamma$-rays in these CSOs may indicate a hallmark of recently restarted jet activity (\citealt{2020ApJ...899..141L}). By studying the $\gamma$-ray emission in radio galaxies with the very long baseline interferometry (VLBI), \cite{2020A&A...641A.152A} suggested that the high-energy emission of radio galaxies is related to the pc-scale radio emission from the inner jet, but the $\gamma$-ray loudness does not be connected with a higher prevalence of Doppler boosting effect (see also \citealt{2019A&A...627A.148A}). The asymmetric jets seen in some CSOs might be evidence for bulk relativistic motion (\citealt{1996ApJ...460..612R}). CSOs with a visibly detected core, especially accompanying variability, also likely indicate the Doppler boosting effect in these core regions (\citealt{2000ApJ...534...90P, 2005ApJ...622..136G}).

CTD 135 (also named as 2234+282) is a CSO located at a redshift of $z=0.795$ (\citealt{1977ApJ...217..358S}). With the very large array (VLA) observation at 1.64 GHz, \cite{1993MNRAS.264..298M} revealed a bright compact component and a weak feature separated about 4.${\arcsec}$9 and reported that CTD 135 is a powerful core-dominated radio source. However, the very long baseline array (VLBA) images at 2.3 GHz show as a single compact component, which is well resolved along the northeast-southwest direction at 8.4 GHz and 15 GHz (\citealt{2016AN....337...65A}). It exhibits a compact triple structure at 15 GHz with a likely core between component-NE and component-SW, and the core contributes the majority of the total flux and the variability (\citealt{2016AN....337...65A}). In this paper, we focus on its properties of $\gamma$-rays observed with the Fermi/LAT in order to reveal the radiation physics and the nature of its symmetric radio structure.

\section{Fermi/LAT Data Reduction and Analysis}

Fermi/LAT is a $\gamma$-ray telescope covering a wide energy range from 20 MeV to more than 300 GeV. It is a powerful tool for monitoring $\gamma$-ray emission from AGNs. Fermi/LAT Pass 8 data, which approximately covers $\sim$11-year (MET 239557417--584235367), is used in our analysis. We only focus on photons within the $15^{\circ}$ region of interest (ROI) centered on the radio position of CTD 135 (R.A. = 339.094, Decl. = 28.483; \citealt{1995AJ....110..880J}). We select the $\gamma$-ray events in the energy range of 0.1--300 GeV, using standard data quality selection criteria ``(DATA\_QUAL$>$0)$\&\&$(LAT\_CONFIG==1)". The maximum zenith angle of $90^{\circ}$ is set to reduce the $\gamma$-ray contamination from the Earth limb. The P8R3\_SOURCE\_V2 set of instrument response function and the publicly available software \textit{fermitools} (ver. 1.0.0) with the binned likelihood analysis method is used in our data analysis.

\begin{figure}
\centering
\includegraphics[angle=0,scale=0.32]{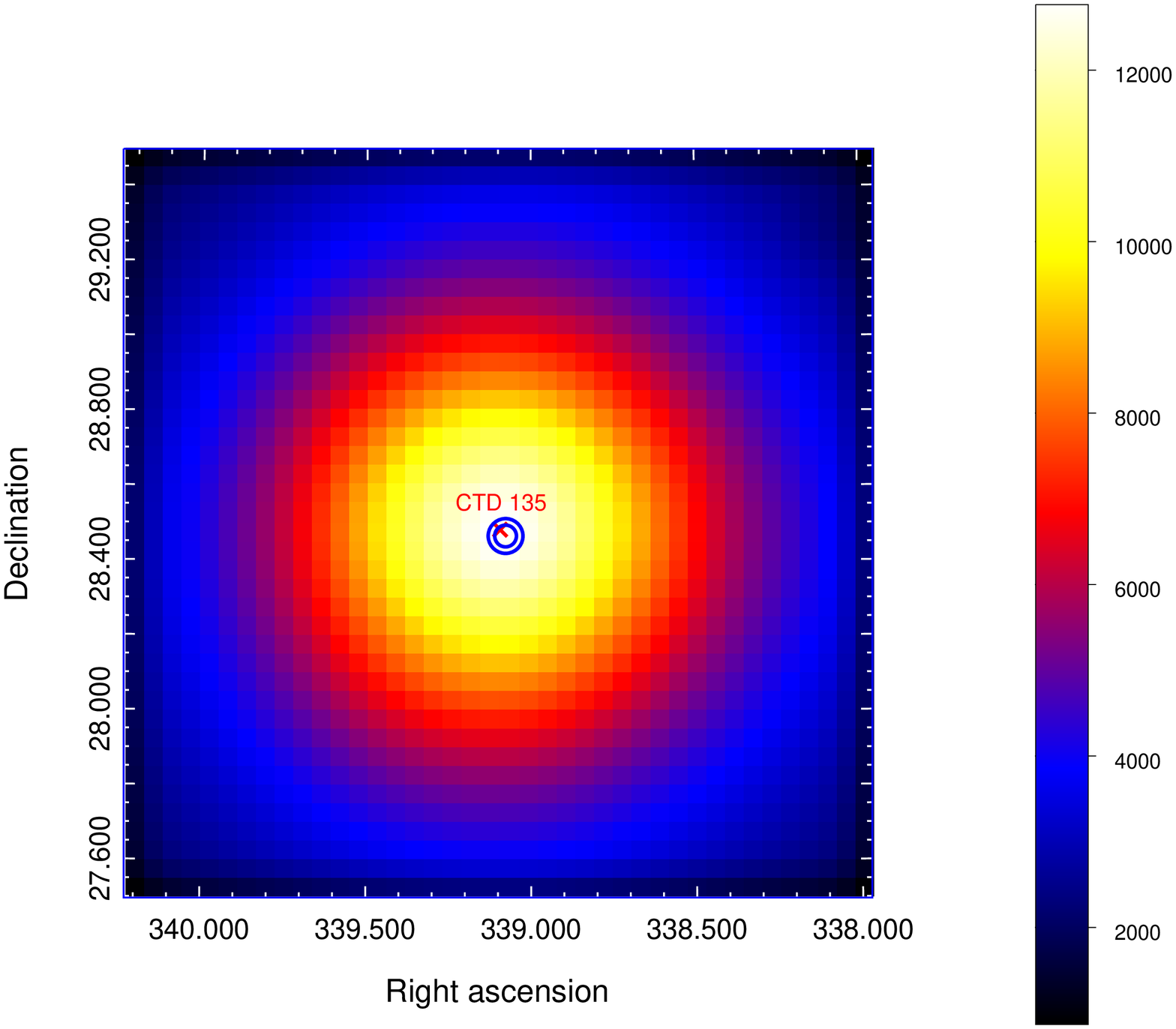}
\caption{$2^{\circ}\times2^{\circ}$ TS map of CTD 135. The red cross represents the radio position of CTD 135. The blue contours represent the 68\% and 95\% confidence levels of the gtfindsrc best-fit position with the $\sim$11-year Fermi/LAT observation data. The map is created with a pixel size of $0.05^{\circ}$.}\label{TSmap}
\end{figure}

We employ the maximum test statistic (TS) to identify a point source from the background, where TS = $2{\rm log}(\frac{\mathcal{L}_{\rm src}}{\mathcal{L}_{\rm null}})$, i.e., the logarithmic ratio of the likelihood for a model with the point source to a model without the point source (\citealt{1996ApJ...461..396M}). A source with a maximum likelihood TS$>25$ is considered as a point source (\citealt{2020ApJS..247...33A}). The background model includes the isotropic emission (i.e., iso\_P8R3\_SOURCE\_V2\_v1.txt) and the diffuse Galactic interstellar emission (i.e., gll\_iem\_v07.fits), as well as all the $\gamma$-ray sources listed in the Fourth Fermi LAT Source Catalog (4FGL, \citealt{2020ApJS..247...33A}). We keep the normalization of the isotropic emission and the diffuse Galactic interstellar emission free. The normalization and spectral parameters of the $\gamma$-ray sources within $10^{\circ}$ in the background model were set to be free. In 4FGL (\citealt{2020ApJS..247...33A}), the $\gamma$-ray source 4FGL J2236.3+2828 was reported to be associated with CTD 135 (see also \citealt{2016AN....337...65A}). Note that the $\sim$11-year Pass 8 data is used in our analysis, but the 4FGL was extracted with the 8-year Fermi/LAT observation data. We firstly make a test of new point sources in the 11-year Fermi/LAT data. After subtracting all the background models, we find that the maximum value in the residual TS map is 7.5, excluding the possibility of new point sources in the 11-year data. Our result is shown in Figure \ref{TSmap}, confirming that CTD 135 is spatially associated with 4FGL J2236.3+2828.

\begin{figure}
\centering
\includegraphics[angle=0,scale=0.32]{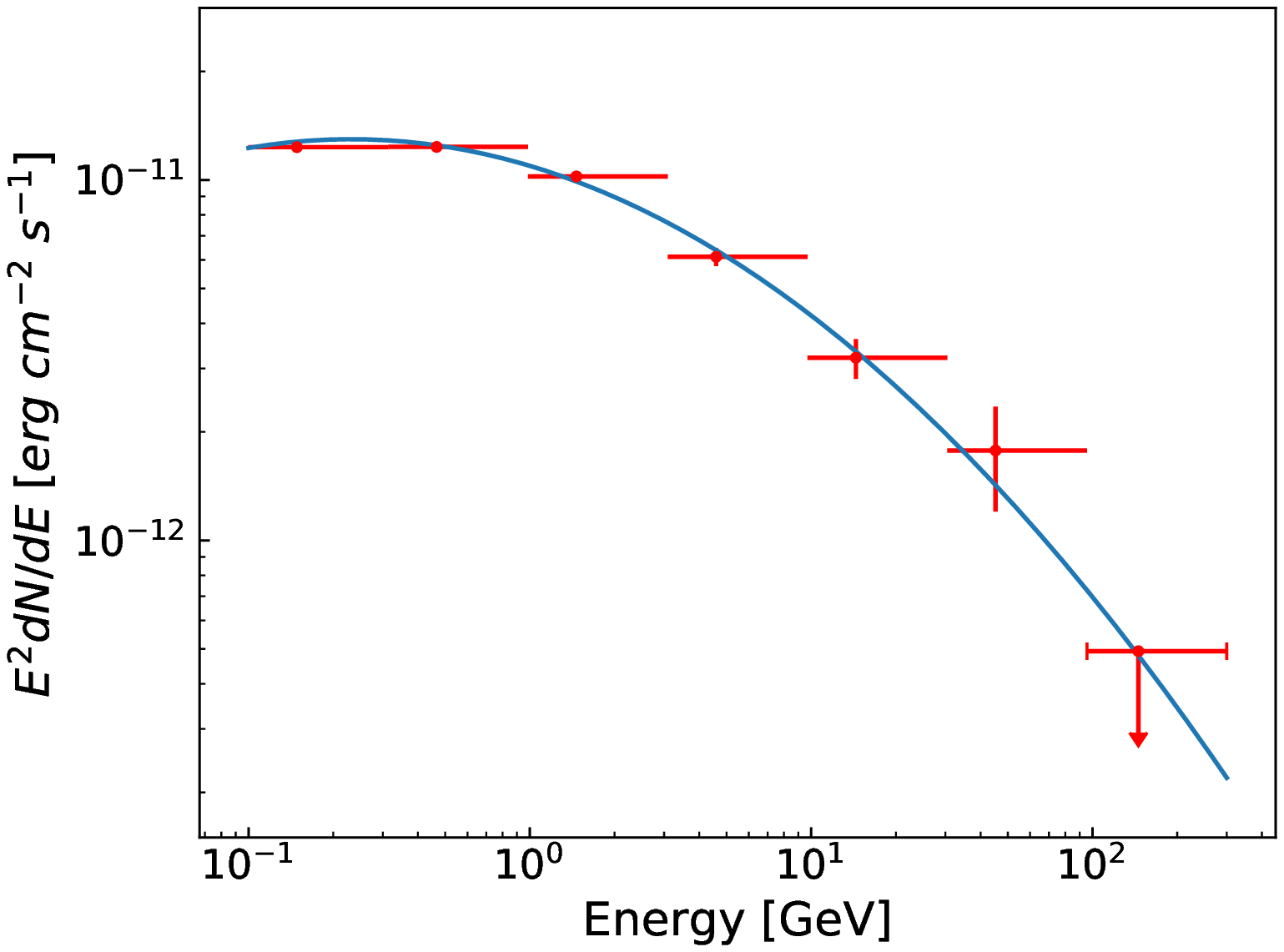}
\includegraphics[angle=0,scale=0.28]{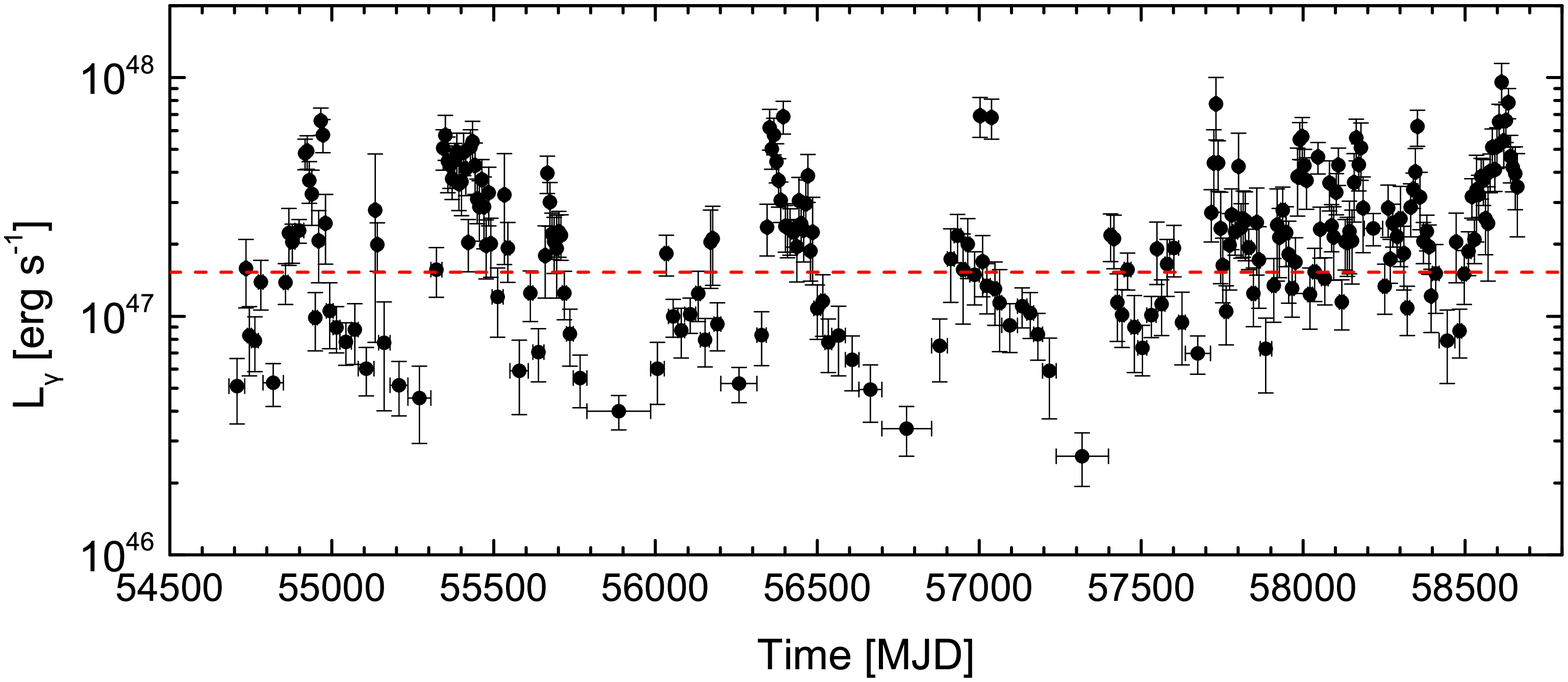}
\caption{Left panel: the $\sim$11-year average spectrum of CTD 135 observed by Fermi/LAT in energy band of 0.1--300 GeV. The blue solid line represents the fitting result with a LogParabole spectral model. Right panel: long-term light curve observed by Fermi/LAT for CTD 135, where a adaptive-binning method based on the constraint of $\rm TS=25$ is used, i.e., the detections of all data points being larger than $5\sigma$. The red horizontal dashed line represents the $\sim$11-year average luminosity of CTD 135.}\label{LAT}
\end{figure}

The ``LogParabole" spectral model is adopted in our analysis for 4FGL J2236.3+2828, i.e.,
\begin{equation}
\frac{dN}{dE} =  N_{0}(\frac{E}{E_{\rm b}})^{-(\Gamma_{\gamma}+\beta{\log}(\frac{E}{E_{\rm b}}))},
\end{equation}
where $E_{\rm b}$ is the reference energy, $\Gamma_{\gamma}$ is the photon spectral index and $\beta$ is the curvature (\citealt{2004A&A...413..489M}). This model is typically used to describe the spectra of blazars in 4FGL. The curvature of a spectrum is evaluated with TS$_{\rm curv}=2{\rm log}(\mathcal{L}_{\rm LP}/\mathcal{L}_{\rm PL})$, where $\mathcal{L}_{\rm LP}$ and $\mathcal{L}_{\rm PL}$ are the likelihoods of hypothesis for testing LogParabole model and power-law model, respectively. In case of TS$_{\rm curv}>9$, the curvature of a spectrum is significant at $\sim3\sigma$ confidence level (\citealt{2020ApJS..247...33A}). The derived TS$_{\rm curv}$ is 90 for 4FGL J2236.3+2828, implying that its spectrum is significantly curved. As shown in Figure \ref{LAT}, the spectral shape of 4FGL J2236.3+2828 can be well fit by the ``LogParabole" spectral model. The best-fit results are $\Gamma_{\gamma}=2.137\pm0.001$, $\beta=0.079\pm0.001$, and  $E_{\rm b}=548.34$ MeV with TS$=12558$. The derived average $\gamma$-ray luminosity is $L_{\gamma}=(1.526\pm0.005)\times10^{47}$ erg s$^{-1}$.

We extract the long-term light curve of CTD 135 by adopting a criterion of TS$>25$ for each time interval and a minimum time-bin of 7 days. Our results are shown in Figure \ref{LAT}. The likelihood-based statistic method is most commonly used to quantify the variability of sources (\citealt{2012ApJS..199...31N, 2019ApJ...884...91P, 2020ApJS..247...33A, 2020ApJ...901..158X}). We evaluate the variability of the $\gamma$-ray emission for CTD 135 with an index of TS$_{\rm var}$ as defined in \cite{2012ApJS..199...31N}, and obtain TS$_{\rm var}=1002$ for CTD 135. Note that if TS$_{\rm var}>25.4$, the confidence level of the variability is $3\sigma$ in a $\chi_{N-1}^{2}(\rm TS_{\rm var})$ distribution with $N-1=9$ degrees of freedom, and $N$ is the number of time-bins. Therefore, CTD 135 is an extremely variable source in the $\gamma$-ray band.

\begin{figure}
\centering
\includegraphics[angle=0,scale=0.35]{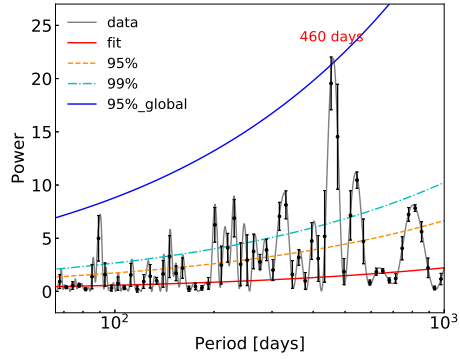}
\caption{LSP of $\gamma$-ray light curve is marked as gray line. The black points with corresponding errors indicate the re-binned data of LSP. The best-fit power spectrum is shown as red solid line. The orange dash line and cyan dot-dashed line represent the single-frequency 95$\%$ and 99$\%$ confidence level lines, respectively. The blue solid line represents the global 95$\%$ false-alarm level of the power-law model.}\label{LSP}
\end{figure}

Inspecting the long-term $\gamma$-ray light curve of CTD 135, one can observe a tentative quasi-periodic oscillation (QPO) signature. We analyze the power density spectrum (PDS) of the $\gamma$-ray light curve and evaluate the QPO signature with the Lomb--Scargle Periodogram (LSP) algorithm (\citealt{1976Ap&SS..39..447L, 1982ApJ...263..835S}; see \citealt{2017ApJ...849...42Z} for more detail). The PDS is presented in figure \ref{LSP}, which shows a visible peak at $\sim$460 days. The highest peak of PDS at $\sim$460 days is beyond the global 95$\%$ false-alarm level, likely indicating that a QPO signature is presented in the $\gamma$-ray light curve of CTD 135.

\section{Broadband SED Constructing and Modeling}

\begin{figure}
\centering
\includegraphics[angle=0,scale=0.4]{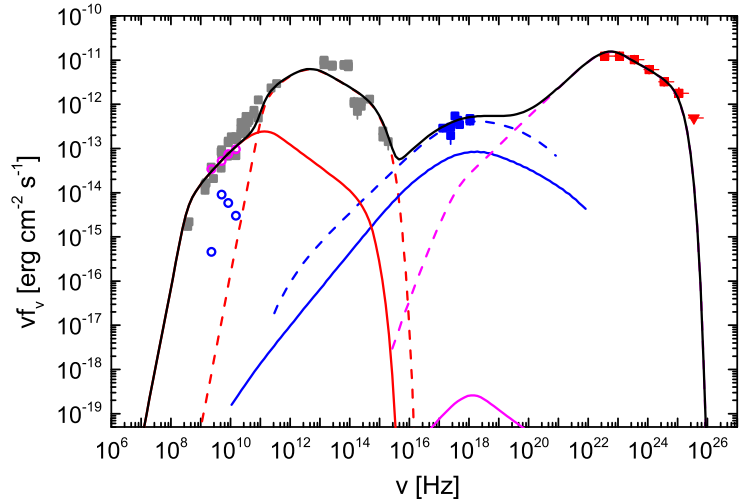}
\caption{Observed SED with the two-zone leptonic radiation model fitting. The gray squares are taken from the NED, the blue squares are taken from the ASDC, and the red squares are the average spectrum of the Fermi/LAT observations. The magenta and blue opened circles indicate the fluxes from component-NE+core and component-SW (taken from \citealt{2016AN....337...65A}), respectively. The black solid line indicates the total emission of the two regions. The colored solid and dashed lines represent the radiations from the extended region and the core region, respectively; red lines for synchrotron, blue lines for SSC, magenta lines for EC. }
\label{SED}
\end{figure}

Using the $\gamma$-ray data derived above together with the other multi-wavelength data collected from the ASI Science Data Center (ASDC) and the NASA/IPAC Extragalactic Database (NED), we compile the broadband spectral energy distribution (SED) of CTD 135, as illustrated in Figure \ref{SED}. The most striking feature of the SED is the triple peaks at the optical, X-ray and GeV $\gamma$-band bands. The broadband SEDs of the $\gamma$-ray emitting compact steep-spectrum sources (CSSs) can be well explained by the two-zone leptonic radiation model (\citealt{2020ApJ...899....2Z}). Hence, we also consider the same model to fit the SED of CTD 135, i.e., the radiations of the relativistic electrons in both compact core and extended region (component-NE and component-SW), including the synchrotron (syn), synchrotron-self-Compton (SSC), and external Compton (EC) scattering of the relativistic electrons. The electron energy distributions in the range of $[\gamma_{\min},\gamma_{\max}]$ for both the core and extended regions are taken as a broken power-law with indices $p_1$ and $p_2$ as well as a break at $\gamma_b$, i.e.,

\begin{equation}
N(\gamma )= N_{0}\left\{ \begin{array}{ll}
\gamma ^{-p_1}  &  \mbox{ $\gamma_{\rm min}\leq\gamma \leq \gamma _{\rm b}$}, \\
\gamma _{\rm b}^{p_2-p_1} \gamma ^{-p_2}  &  \mbox{ $\gamma _{\rm b} <\gamma <\gamma _{\rm max} $.}
\end{array}
\right.
\end{equation}

For the core region, the radiation region is assumed as a homogenous sphere with radius $R$ and magnetic field strength $B$, where $R$ is estimated with $R=\delta c \Delta t/(1+z)$. Since the observed time-scale of $\gamma$-ray flux variation is smaller than 7 days (as displayed in Figure \ref{LAT}), we take $\Delta t=7$ days. The brightness temperature and variability of the radiations from the core region indicate its Doppler boosting effect (\citealt{2009A&A...494..527H, 2016AN....337...65A, 2018ApJ...866..137L}). We thus consider the relativistic motion of the core region with the bulk Lorentz factor $\Gamma$ and the Doppler factor $\delta$. Since the $\gamma$-ray luminosity of CTD 135 is bright and variable as blazars, we fit the SED by assuming $\delta=\Gamma$ as done for blazars (e.g., \citealt{2012ApJ...752..157Z, 2015ApJ...807...51Z}). The photon field from the outer region of broad line regions (BLRs) is used to calculate the EC process\footnote{Since the radiation region is far away from the black hole and the accretion disk emission is not important anymore, we thus do not consider the accretion disk radiation as the external photon field (e.g., \citealt{2009MNRAS.397..985G}).}. The minimum energy of the radiation electrons is fixed as $\gamma_{\min}=1$ and the maximum energy $\gamma_{\max}$ is poorly constrained by the last observation point in the $\gamma$-ray band. Spectral indices $p_1$ and $p_2$ are taken the values derived from the observed SEDs.

For the extended region, the radiation region is also assumed as a homogenous sphere with an angular radius of 5 mas (approximate distance between component-NE and component-SW in  \citealt{2016AN....337...65A}). The derived average hot-spot advance speed between component-NE and component-SW is $\sim0.3c$ (\citealt{2016AN....337...65A}). The symmetric radio structure may also imply a large viewing angle to the extended region, we thus do not consider the relativistic motion and the Doppler boosting effect. The contribution of the inverse Compton scattering of cosmic microwave background photons (IC/CMB) by the relativistic electrons is considered. The SED modeling for the extended region is under the equipartition condition, i.e., the energy densities of the magnetic fields ($U_{B}$) and electrons ($U_{\rm e}$) are equal. We also set $\gamma_{\min}=1$ and $\gamma_{\max}=50*\gamma_{\rm b}$. $p_1$ is fixed as the value derived by fitting the radio spectrum and $p_2$ is fixed as $p_2=4$.

There are not enough simultaneous observational data to constrain the parameters as done for blazars in our previous works (e.g., \citealt{2012ApJ...752..157Z, 2015ApJ...807...51Z}), so the goodness of SED fitting is assessed visually. The synchrotron-self-absorption, the Klein--Nishina effect, and the absorption of high-energy $\gamma$-ray photons by extragalactic background light (\citealt{2008A&A...487..837F}) are also taken into account. As illustrated in Figure \ref{SED}, the model can well represent the SED. We obtain $R=1.1\times10^{17}$ cm, $B=0.97$ G, $\delta=\Gamma=10.8$, $\gamma_{\rm b}=373$, $\gamma_{\rm max}=5500$, $N_{0}=286.4$ cm$^{-3}$, $p_1=1.7$, $p_2=3.76$, $U_{\rm BLR}=1.1\times10^{-3}$ erg cm$^{-1}$ (energy density of the outer region for BLR) for the core region; $R=37.51$ pc, $B_{\rm eq}=8.57$ mG, $\gamma_{\rm b}=2200$, $N_{0}=0.012$ cm$^{-3}$, and $p_1=1.35$ for the extended region, respectively. It is found that the $\gamma$-rays are contributed by the EC/BLR process, the X-ray emission is dominated by the SSC process of the core region, and the IR-optical bump is attributed to the synchrotron emission of the electrons in the core region. Although the electrons in the extended region are more energetic than that in the core region, their synchrotron radiations peak at $10^{10}$ Hz since the magnetic field strength of the extended region is much lower than that of the core region. The radiations from the SSC process of both the core and extended regions peak at a similar frequency range. And the emission of the IC/CMB process from the extended region also peaks at the same range, but its flux is extremely low.

\begin{figure}
\centering
\includegraphics[angle=0,scale=0.32]{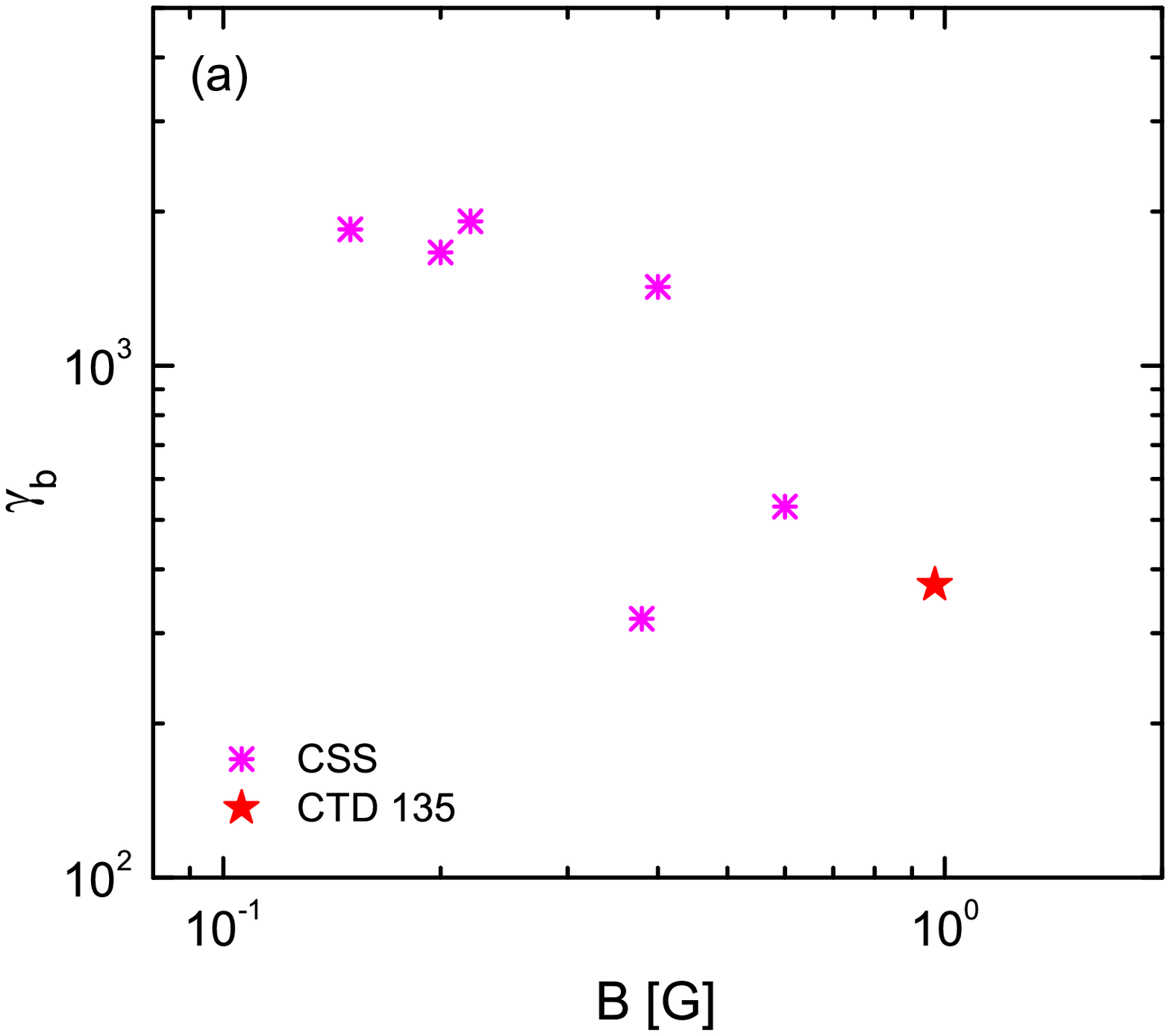}
\includegraphics[angle=0,scale=0.32]{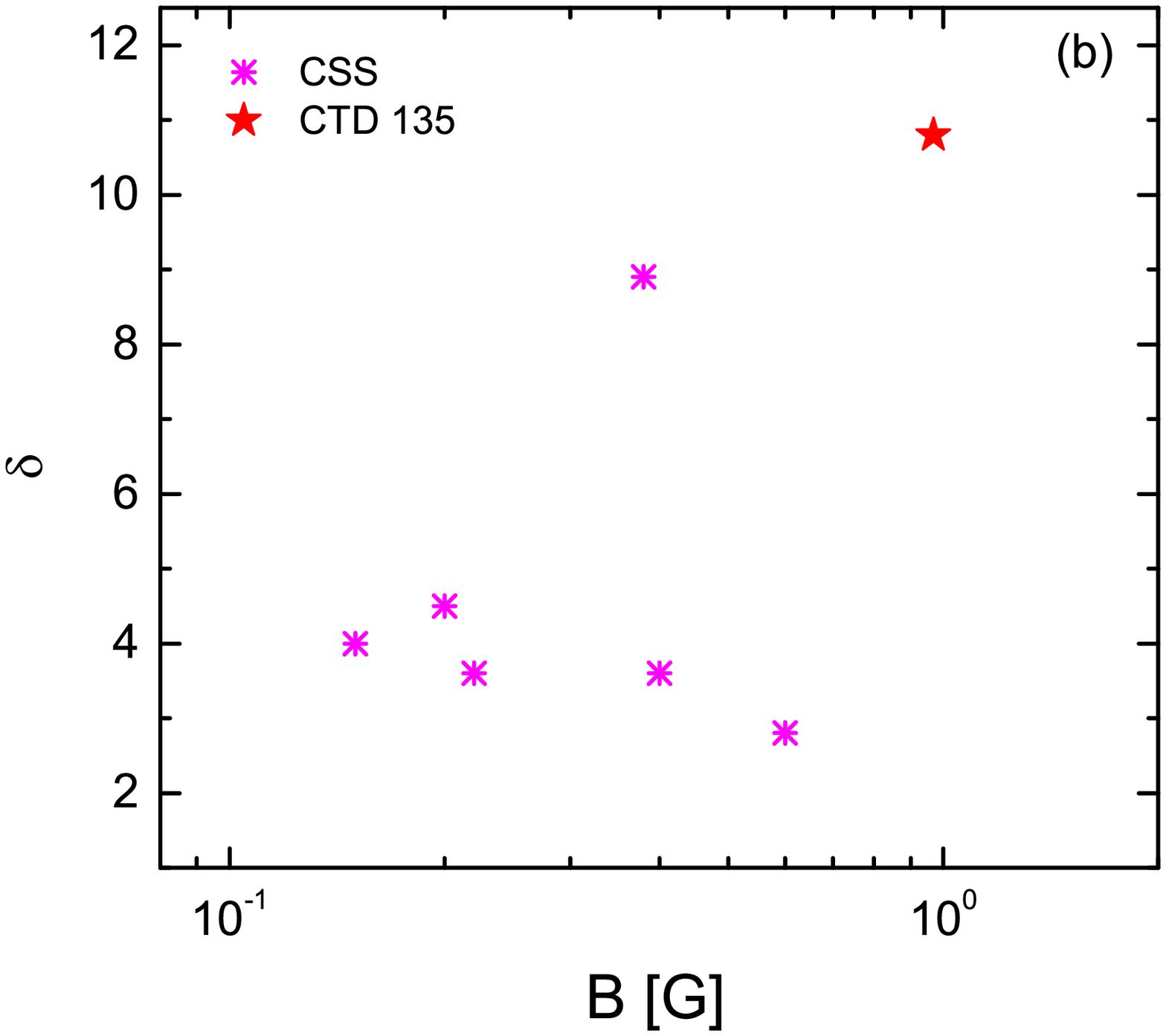}
\caption{$\gamma_{\rm b}$ and $\delta$ as a function of $B$. The data of six $\gamma$-ray emitting CSSs are taken from \cite{2020ApJ...899....2Z}.}
\label{gamb_B_delta}
\end{figure}

We obtain $\delta=10.8$ for the core region, which is larger than the variability Doppler factor of $\delta_{\rm var}=6$ inferred by the flux-density variations at the radio band in \cite{2009A&A...494..527H}, but smaller than the reported value of $\delta_{\rm var}=20.19$ in \cite{2018ApJ...866..137L}. Both CSOs and CSSs belong to young radio-loud AGNs. As displayed in Figure \ref{gamb_B_delta}, the core region of CTD 135 has the larger $B$ and $\delta$, on average smaller $\gamma_{\rm b}$ than those of $\gamma$-ray emitting CSSs (the data of CSSs to see \citealt{2020ApJ...899....2Z}), indicating that the stronger radiation cooling for the electrons by the magnetic field and EC process in CTD 135 than in those CSSs. Similar to the $\gamma$-ray emitting CSSs, $p_1=1.7$ is smaller than the expected value of 2 from the shock acceleration mechanism (e.g., \citealt{2000ApJ...542..235K, 2001MNRAS.328..393A, 2005ApJ...621..313V}). The flatter power-law particle spectrum may be due to the magnetic reconnection acceleration process (\citealt{2001ApJ...562L..63Z, 2014ApJ...783L..21S, 2015ApJ...806..167G, 2016RAA....16..170Z}). For the extended region, the derived equipartition magnetic filed of $B_{\rm eq}=8.57$ mG is comparable to the typical range of 1--10 mG measured in CSOs (\citealt{2006A&A...450..959O, 2020ApJ...899..141L}).

\section{Discussion}

\subsection{Powers of the Core Region in Comparison with $\gamma$-ray Emitting CSOs and CSSs as well as Blazars}

\begin{figure}
\centering
\includegraphics[angle=0,scale=0.32]{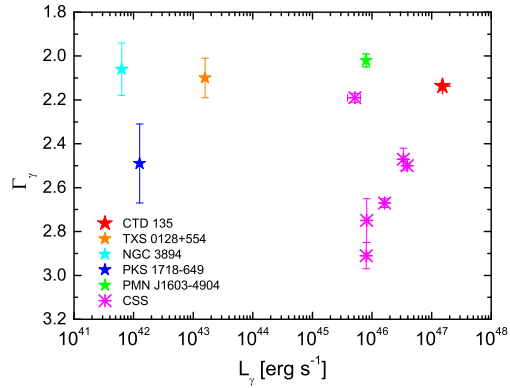}
\caption{$\Gamma_{\gamma}$ against $L_{\gamma}$, where $\Gamma_{\gamma}$ and $L_{\gamma}$ of CTD 135 are the $\sim$11-year average photon spectral index and $\gamma$-ray luminosity, respectively, observed by Fermi/LAT. The data of six $\gamma$-ray emitting CSSs and other four $\gamma$-ray emitting CSOs are taken from \cite{2020ApJ...899....2Z} and \cite{2020ApJS..247...33A}, respectively.}
\label{Gamma-L}
\end{figure}

CSOs may evolve first into CSSs and then into FR II radio galaxies (\citealt{1996ApJ...460..634R}). As displayed in Figure \ref{Gamma-L}, CTD 135 is the brightest one among the $\gamma$-ray selected CSOs and CSSs. The $\sim$11-year average luminosity of CTD 135 is $L_{\gamma}\sim10^{47}$ erg s$^{-1}$, and $L_{\gamma}$ is almost close to $10^{48}$ erg s$^{-1}$ in some time-bins as shown in Figure \ref{LAT}. As reported in Abdollahi et al. (2020; see also \citealt{2020ApJ...899..141L}), NGC 3894 is the lowest luminosity CSO with $L_{\gamma}<10^{42}$ erg s$^{-1}$ and a flat photon index of $2.06\pm0.12$; PKS 1718--649 also has a low luminosity ($L_{\gamma}\sim10^{42.1}$ erg s$^{-1}$), but has a steeper photon index of $2.49\pm0.18$ than NGC 3894; both TXS 0128+554 and PMN 1603--4904 have the flat photon index similar to NGC 3894 with the higher luminosity ($10^{43}<L_{\gamma}<10^{46}$ erg s$^{-1}$). Different from CTD 135, the four $\gamma$-ray emitting CSOs do not show fast and obvious variabilities at the $\gamma$-ray band. $L_{\gamma}$ of CTD 135 is similar to some bright flat-spectrum radio quasars (FSRQs, a sub-class of blazars). The derived values of $B$, $\delta$, and $\gamma_{\rm b}$ are also similar to those of blazars (\citealt{2012ApJ...752..157Z, 2015ApJ...807...51Z}).

On the basis of the SED fitting parameters, we calculate the jet powers in case of the e$^{\pm}$ pair jet for the core region of CTD 135. The powers of radiation electrons ($P_{\rm e}$) and magnetic fields ($P_{B}$) are calculated by $P_{\rm i}=\pi R^{2}{\Gamma}^{2}cU_{\rm i}$, where $U_{\rm i}$ is $U_{\rm e}$ or $U_{B}$. The radiation power is $P_{\rm r} = \pi R^{2}\Gamma^{2}cU_{\rm r} = L_{\rm bol}\Gamma^{2}/4\delta^4$, where $L_{\rm bol}$ is the non-thermal radiation luminosity of the core region. We obtain $P_{\rm e}=6.03\times 10^{44}$ erg s$^{-1}$, $P_{B}=4.89\times10^{45}$ erg s$^{-1}$, $P_{\rm r}=7.29\times10^{44}$ erg s$^{-1}$, and $P^{e^{\pm}}_{\rm jet}=P_{\rm e}+P_{B}+P_{\rm r}=6.23\times10^{45}$ erg s$^{-1}$. We also obtain $P_{\rm r}/P_{\rm jet}^{e^{\pm}}=0.12$ and $P_{B}/P_{\rm jet}^{e^{\pm}}=0.79$, indicating that the core region of CTD 135 may be highly magnetized with high radiation efficiency, similar to FSRQs and narrow-line Seyfert 1 galaxies (NLS1s) as well as $\gamma$-ray emitting CSSs (\citealt{2014ApJ...788..104Z, 2015ApJ...807...51Z, 2020ApJ...899....2Z, 2015ApJ...798...43S}).

\begin{figure}
\centering
\includegraphics[angle=0,scale=0.35]{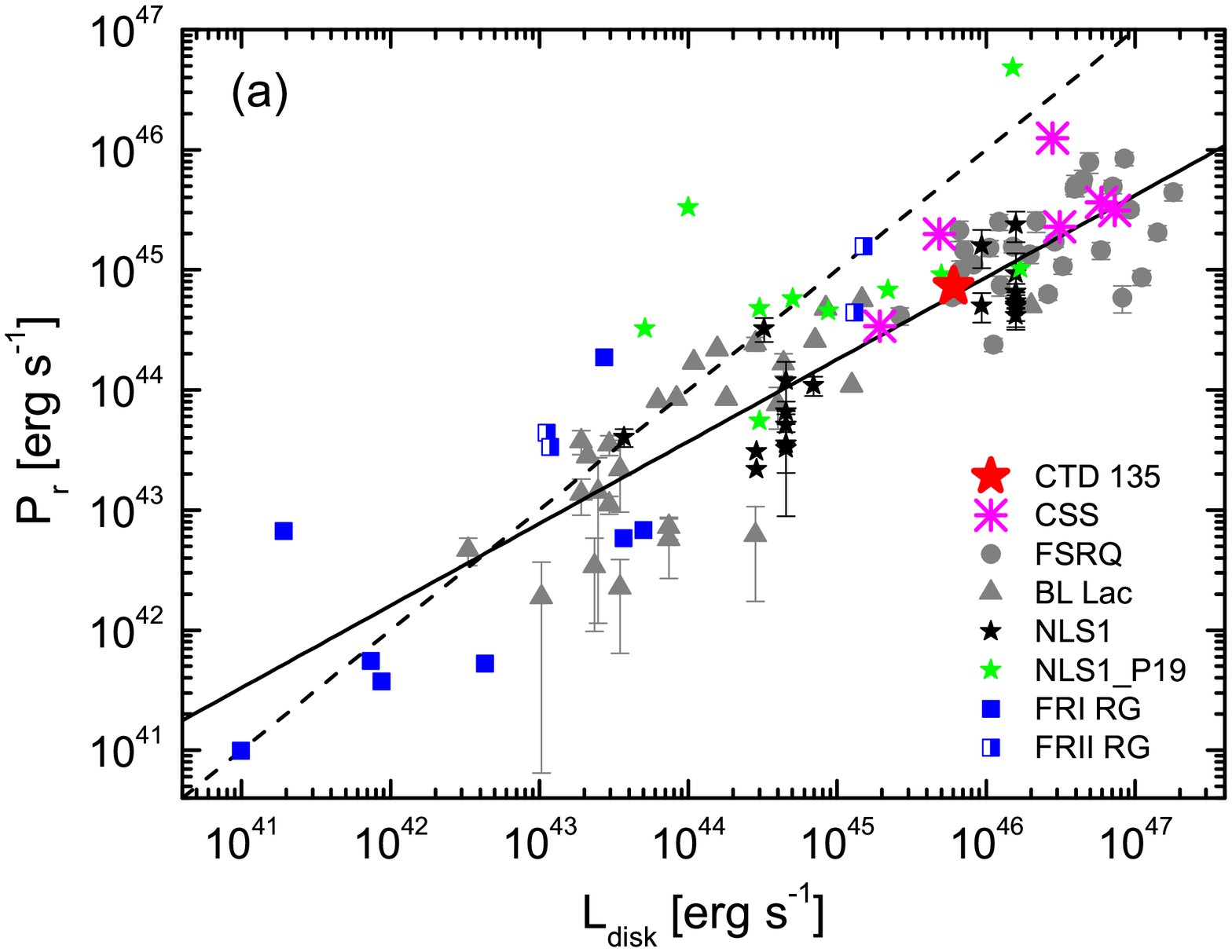}
\includegraphics[angle=0,scale=0.35]{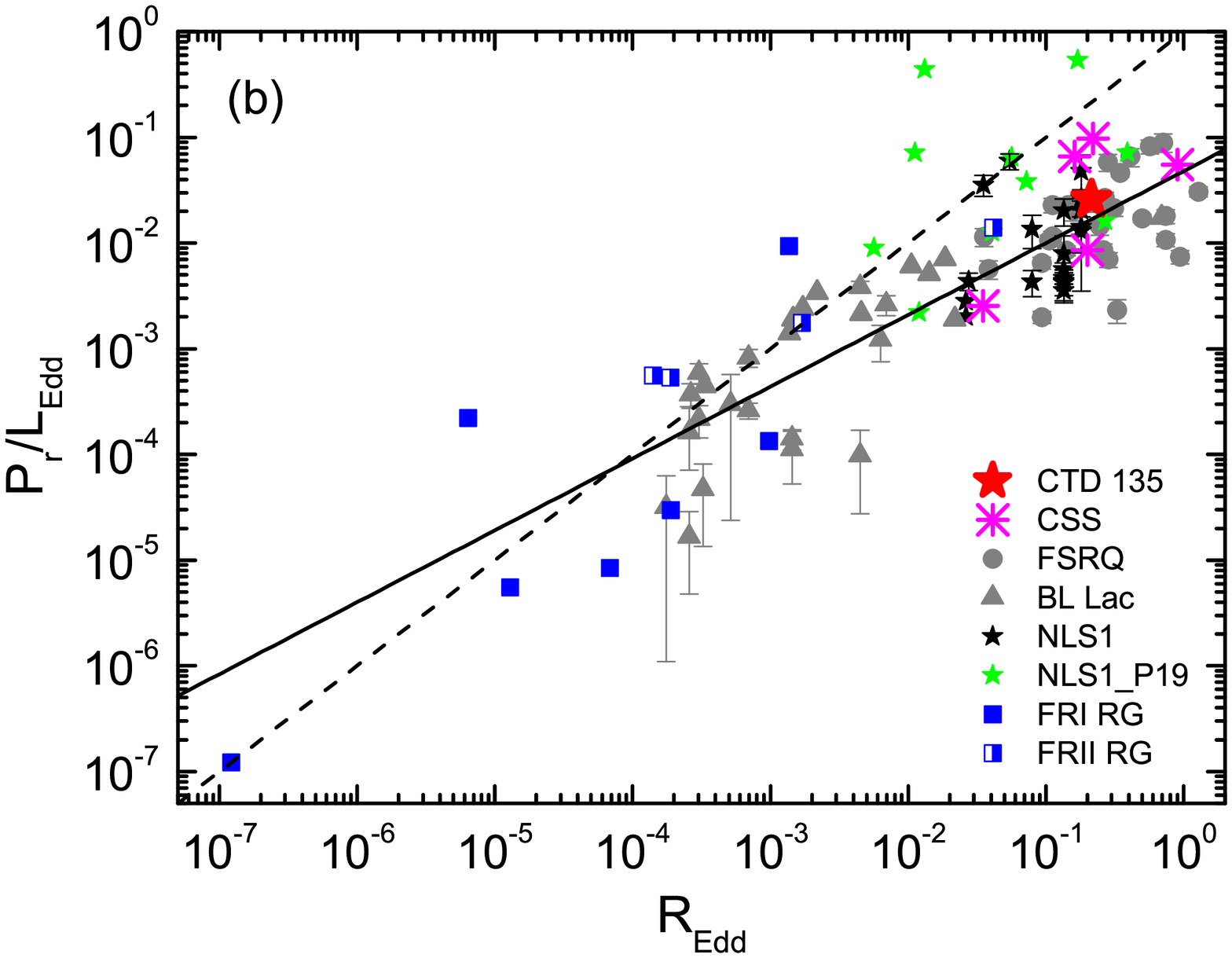}
\caption{$P_{\rm r}$ as a function of $L_{\rm disk}$ together with their relations in units of Eddington luminosity for CTD 135. The dashed lines indicate the equality line while the solid lines are the linear regression fits for other detected $\gamma$-ray emitting AGNs (including blazars, NLS1s, radio galaxies, and CSSs, taken from \citealt{2020ApJ...899....2Z}).}
\label{Pr-Ld}
\end{figure}

As shown in Figure \ref{SED}, no apparent thermal-radiation component from the accretion disk is observed in the broadband SED of CTD 135. If the disk luminosity ($L_{\rm disk}$) of CTD 135 is estimated by $L_{\rm disk}=10*L_{\rm BLR}$, where $L_{\rm BLR}$ is the BLR luminosity and taken from \cite{1997MNRAS.286..415C}, we obtain $L_{\rm disk}\sim6.03\times10^{45}$ erg s$^{-1}$. The central black hole mass is $M_{\rm BH}=10^{8.35}M_{\bigodot}$ (\citealt{2012ApJ...748...49S}), hence the derived Eddington-ratio ($R_{\rm Edd}$) of CTD 135 is $R_{\rm Edd}=L_{\rm disk}/L_{\rm Edd}\sim0.21$, where $L_{\rm Edd}$ is the Eddington luminosity. We plot $P_{\rm r}$ against $L_{\rm disk}$ together with their relations in units of Eddington luminosity for CTD 135 and other $\gamma$-ray emitting AGNs (the data to see \citealt{2020ApJ...899....2Z}), as illustrated in Figure \ref{Pr-Ld}. We find that CTD 135, similar to $\gamma$-ray emitting CSSs, also obeys the linear regression fit line of other subclasses of $\gamma$-ray emitting AGNs in \cite{2020ApJ...899....2Z}, likely implying that its jet radiation is also driven by the Eddington ratio and $R_{\rm Edd}$ would be a key physical driver for the unification scheme of AGN jet radiation.

\subsection{an Aligned Jet and the Symmetric Radio Structure of CTD 135}

Different from other $\gamma$-ray emitting CSOs and CSSs (\citealt{2014A&A...562A...4M, 2016ApJ...821L..31M, 2020A&A...635A.185P, 2020ApJ...899..141L, 2020ApJ...899....2Z}), significant variability and high luminosity in $\gamma$-ray band are observed for CTD 135 as blazars. The derived variability index is $\rm TS_{\rm var}=1002$ for CTD 135, which is similar to the values of some typical BL Lacs (another kind of blazars), such as $\rm TS_{\rm var}=1349$ for Mrk 421 and $\rm TS_{\rm var}=543$ for Mrk 501 (\citealt{2020ApJS..247...33A}). It is also interesting that a QPO signature is detected in the $\gamma$-ray light curve of CTD 135, which is generally observed in blazars (\citealt{2014ApJ...793L...1S, 2017A&A...600A.132S, 2015ApJ...810...14A, 2018ApJ...867...53Y, 2018NatCo...9.4599Z, 2020ApJ...891..163Z}). Generally, the QPO signature is thought to be due to a super-massive black hole binary system (e.g., \citealt{1988ApJ...325..628S, 1996ApJ...460..207L, 2015Natur.518...74G}).  The precession (e.g., \citealt{2003MNRAS.341..405S, 2013MNRAS.428..280C}) or helical structure (e.g., \citealt{1993ApJ...411...89C, 1999A&A...347...30V, 2004ApJ...617..123N, 2021ApJ...906..105C}) of jets in blazars can also lead to the QPO signal. \cite{2017A&A...600A.132S} suggested that the QPOs in these jet-dominant sources are likely owing to the instability or structure of a jet and the oscillation is amplified by the Doppler boosting effect. All these $\gamma$-ray emission properties favor the idea that the core emission of CTD 135 has significant Doppler boosting effect like a blazar jet.

As described in Section 3, the broadband SED of CTD 135 can be well fitted with the leptonic model by assuming a small viewing angle ($\theta$), being similar to blazars. We assumed $\delta=\Gamma$ during the SED fitting for CTD 135, i.e., the viewing angle ($\theta$) being equal to the opening angle ($1/\Gamma$) of jet (e.g., \citealt{2021ApJ...906..105C}), and thus it is $\theta\sim5.3^\circ$ for $\delta=\Gamma=10.8$. Using the VLBI measurements of the core angular dimension and radio flux and comparing the predicted and observed X-ray flux in the SSC model frame, \cite{1993ApJ...407...65G} estimated the Doppler factor of $\delta=5.2$ and then obtained $\theta<11.1^\circ$ for CTD 135. Using the flux-density variations at radio band, \cite{2018ApJ...866..137L} inferred the variability Doppler factor of $\delta_{\rm var}=20.19$ and $\theta<2.84^\circ$ (corresponding to $\Gamma\sim10.38$). These estimates are generally consistent with the values of blazar jets and imply a small viewing angle of jet in CTD 135. However, the symmetric radio structure of CSOs is thought to be due to a misaligned jet to the observers (\citealt{1980ApJ...236...89P, 1994ApJ...432L..87W, 1996ApJ...460..634R}). We thus propose that the symmetric radio structure of CTD 135 consists of a relativistic jet surrounding by a bubble similar to the Fermi bubbles in the Galaxy (\citealt{2010ApJ...724.1044S}). The $\gamma$-rays of CTD 135 are dominated by the relativistic jet radiation as blazars and the symmetric radio structure (symmetric lobes on both sides of the core) is attributed to bubbles.

The Fermi bubbles may be created by a recent AGN jet activity (\citealt{2012ApJ...756..181G}) or the remnants of a large-scale wide-angle outflow from the central supermassive black hole (\citealt{2011MNRAS.415L..21Z}). Recently \cite{2020ApJ...899..141L} reported that the lack of compact, inverted spectrum hotspots and an emission gap between the bright inner jet and outer radio lobe structure for $\gamma$-ray emitting CSO TXS 0128+554 is due to the episodic jet activity. The episodic jet activity was theoretically suggested for young radio sources (e.g., \citealt{1997ApJ...487L.135R, 2009ApJ...698..840C}). We thus speculate that the pc-scale lobes of CTD 135 are remnants from the previous period of jet activity and the observed $\gamma$-rays are associated with the inner jet that was launched more recently, similar to TXS 0128+554. And the re-launched jet of CTD 135 has a small viewing angle like blazar jet, which leads to the strong Doppler boosting effect, the significant variability, and the high $\gamma$-ray luminosity. Due to the episodic/short-lived jet activity, some weak CSOs like TXS 0128+554 may never grow to large sizes (\citealt{2020ApJ...899..141L}) while some strong CSOs like CTD 135 may evolve into large-scale radio galaxies.

\section{Summary}

In this paper, we comprehensively analyzed the $\sim$11-year Fermi/LAT observation data of CTD 135, confirming that it is spatially associated with 4FGL J2236.3+2828. CTD 135 has a high $\gamma$-ray luminosity similar to FSRQs. The average spectrum of CTD 135 in the LAT energy band is very curved and needs a LogParabole spectral model to fit. The variability index of $\rm TS_{\rm var}=1002$ for CTD 135 indicates that its $\gamma$-ray fluxes are obviously variable. And a QPO signature with a periodicity of $\sim$460 days at a confidence level of $\sim$95\% is presented in the $\gamma$-ray light curve of CTD 135. The high luminosity and obvious variability as well as the QPO signature in the $\gamma$-ray band demonstrate the strong Doppler boosting effect in CTD 135. The strong Doppler boosting effect implies a small viewing angle of the core jet for CTD 135, which is not coincident with its symmetric radio structure. We proposed that an episodic jet exists in the core region of CTD 135. Its pc-scale lobes are remnants from the previous period of jet activity similar to the Fermi bubbles in the Galaxy while its $\gamma$-rays originate from a relaunched relativistic jet.

We also constructed the broadband SED of CTD 135 using the derived $\gamma$-ray spectrum together with the data from the ASDC and NED. It can be well explained with the two-zone leptonic model and the $\gamma$-ray emission with obvious variability should be from the radiations of the core jet. The derived values of $B$, $\gamma_{\rm b}$, and $\delta$ for the core region are similar to those of blazars. On the base of SED fitting parameters, we calculated $P_{\rm r}$, $P_{\rm e}$, $P_{B}$, and $P_{\rm jet}^{e^{\pm}}$ of the core region for CTD 135. Similar to $\gamma$-ray emitting CSSs, CTD 135 also has the large values of $P_{\rm r}/P_{\rm jet}^{e^{\pm}}$ and $P_{B}/P_{\rm jet}^{e^{\pm}}$, indicating that its core region may be highly magnetized with high radiation efficiency. We estimated the disk luminosity in CTD 135 with $L_{\rm disk}=10*L_{\rm BLR}$, and then plotted $P_{\rm r}$ against $L_{\rm disk}$ together with their relation in units of Eddington luminosity. We found that CTD 135 also follows the linear regression fit line of other subclasses of $\gamma$-ray emitting AGNs in \cite{2020ApJ...899....2Z}. Hence, we proposed that the jet radiation of CTD 135 is also driven by the Eddington ratio, similar to other $\gamma$-ray emitting AGNs.

\begin{acknowledgements}
This work is supported by the National Natural Science Foundation of China (grants 12022305, 11973050, 11863007, U1738136, U1731239, 11851304, and 11533003), and Guangxi Science Foundation (grants 2017AD22006, 2019AC20334, and 2018GXNSFGA281007).
\end{acknowledgements}

\clearpage

\end{document}